\documentclass[twocolumn,pre,nofootinbib,superscriptaddress]{revtex4}

\newcommand{\beq}{\begin{eqnarray}}
\newcommand{\eeq}{\end{eqnarray}}

\usepackage{graphicx,amsmath,tabularx}
\usepackage{color} 

\begin{document}

\title{Finite-size scaling above the upper critical dimension}
\author{Matthew Wittmann}
\affiliation{Department of Physics, University of California, Santa Cruz, California 95064, USA}
\author{A.~P.~Young}
\affiliation{Department of Physics, University of California, Santa Cruz, California 95064, USA}
\affiliation{Max Planck Institute for the Physics of Complex Systems,
N\"othnitzer Strasse, 01187 Dresden, Germany}

\date{\today}

\begin{abstract}
We present a unified view of finite-size scaling (FSS) in dimension $d$ above
the upper critical dimension, for both free and periodic boundary conditions.
We find that the modified FSS proposed some time ago to allow for violation of
hyperscaling due to a dangerous irrelevant variable, applies only to ${\bf
k}=0$ fluctuations, and so there is only a single exponent $\eta$ describing
power-law decay of correlations at criticality, in contrast to
recent claims. With free boundary conditions
the finite-size ``shift'' is greater than the rounding. Nonetheless, using
$T-T_L$, where $T_L$ is the finite-size pseudocritical temperature, rather
than $T-T_c$, as the scaling variable, the data does collapse on to a scaling
form which includes the behavior both at $T_L$, where the susceptibility
$\chi$ diverges like $L^{d/2}$ and at the bulk $T_c$ where it diverges like
$L^2$.  These claims are supported by large-scale simulations on the
5-dimensional Ising model.
\end{abstract}

\maketitle

\section{Introduction}
\label{sec:intro}

The method of finite size scaling
(FSS)~\cite{fisher:71,fisher:72b,binder:01}
has been success fully applied to the
analysis of the results of many numerical simulations. The main ingredient is
the assumption that finite size corrections only involve the ratio of the system
size $L$ to the bulk (i.e.\ infinite system size) correlation length $\xi$.
The latter diverges as $T$ approaches the transition temperature $T_c$ like
$\xi \propto (T-T_c)^{-\nu}$ where 
$\nu$ is the correlation length exponent.

While this assumption is undoubtedly correct in dimensions below the upper
critical dimension $d_u$, equal to four for most systems, the situation is,
surprisingly, more complicated for $d > d_u$, even though the critical
exponents are given by their mean field values in this region. The reason is that a
``dangerous irrelevant''variable causes scaling functions to have
additional singularities.  While the nature of FSS above the upper critical
dimension has been clarified for ${\bf k} = 0$ fluctuations in
systems with periodic boundary conditions, the situation in models with free
boundary conditions, and for ${\bf k} \ne 0$ fluctuations for both boundary
conditions, seems confused. The purpose of the work presented here is to clarify
these questions and present a simple, unified, picture of FSS above the upper
critical dimension. 

According to \textit{standard} FSS, valid for $d < d_u$, a susceptibility $\chi$ which
diverges in the bulk like $(T-T_c)^{-\gamma}$ for $T \to T_c$, has a FSS form
\begin{equation}
\chi(L, T) = L^{\gamma y_T} \, \overline{X}\left(L^{y_T}\, (T-T_c)\right)
\, ,
\label{chi_standard}
\end{equation}
where $y_T$ is the
thermal exponent in the renormalization
group sense and is related to the correlation length exponent by
\begin{equation}
y_T = {1 \over \nu} \, .
\label{yT}
\end{equation}
The argument of the scaling function $\overline{X}$ is proportional to 
$(L/\xi)^{1/\nu}$ so Eq.~\eqref{chi_standard} implements the basic
FSS assumption, stated above,
that finite-size effects depend on the ratio $L/\xi$~\cite{power}.
Above $T_c$ and for large $L$, finite-size effects
disappear so we must recover the bulk result, which requires $\overline{X}(x)
\propto x^{-\gamma}$ for $x \to \infty$.

Finite-size scaling is particularly simple for dimensionless (more generally
scale-invariant) quantities for which the exponent $\gamma$ above is zero.
An example is the dimensionless ratio of the moments of the order parameter
proposed by Binder~\cite{binder:81}. 
The Binder ratio, $g$, defined in Eq.~\eqref{binder} below,
has the standard FSS form 
\begin{equation}
g(L, T) = \overline{g}\left(L^{y_T}\, (T-T_c)\right)
\, .
\label{g_standard}
\end{equation}
One sees that the data is independent of size at $T_c$ so data for different
sizes intersect there, which provides a very convenient way of locating $T_c$.
Furthermore, the scaling functions $\overline{X}(x)$ and $\overline{g}(x)$ are
predicted to be universal (apart from a non-universal metric factor multiplying
the argument $x$, and a non-universal factor multiplying
the prefactor $L^{\gamma y_T}$ in
Eq.~\eqref{chi_standard}), so the value of $g$ at $T_c$ is predicted to be universal.

The purpose of the present work is to discuss how Eqs.~\eqref{chi_standard}
and \eqref{g_standard} are modified for $d > d_u = 4$. First of all we note that,
in this region, we have mean-field exponents whose values are $\gamma = 1,
y_T = 1/\nu = 2$ so naively we would have
\begin{subequations}
\begin{align}
\chi(L, T) &= L^2 \, \overline{X}\left(L^2\, (T-T_c)\right),
\label{chi_naive} \\
g(L, T) &= \overline{g}\left(L^2\, (T-T_c)\right) \, .
\end{align}
\label{naive}
\end{subequations}
As discussed above, the power $2$ in these equations is the value of
the thermal exponent $y_T \ (=1/\nu)$
in the mean-field region.
For periodic boundary conditions and, implicitly, for ${\bf k} = 0$
fluctuations, Binder et al.\ \cite{binder:85} showed that one should not use
the thermal exponent $y_T$ but rather modify 
Eq.~\eqref{naive} to
\begin{subequations}
\begin{align}
\chi(L, T) &= L^{y_T^\star} \, \overline{X}\left(L^{y_T^\star}\, (T-T_c)\right),
\label{chi_BNPY}\\
g(L, T) &= \overline{g}\left(L^{y_T^\star}\,
(T-T_c)\right)  \quad \text{(periodic, {\bf k}=0)},
\label{g_BNPY}
\end{align}
\label{BNPY}
\end{subequations}
where
\begin{equation}
y_T^\star = d / 2\, .
\label{yTstar}
\end{equation}
Since $d > 4$, we have $y_T^\star > y_T \ (=2)$.

The universal value of $\overline{g}$ at $T_c$ was computed by Br\'ezin and
Zinn-Justin~\cite{brezin:85} who showed it to be simply that obtained by including
\textit{only} the ${\bf k} = 0$ mode (with $T_c$ adjusted to the
correct value).  An extensive set of
works, see for
example,~\cite{luijten:96,parisi:96b,blote:97,luijten:99,binder:01} and
references therein,
have shown the validity of
Eq.~\eqref{BNPY}, though it required large system sizes, good statistics, and
an appreciation that \textit{corrections} to FSS (which occur if the sizes are
not big enough) are quite large and slowly decaying, to confirm the predicted,
universal value of the Binder ratio at $T_c$.

Equation \eqref{BNPY} is for periodic boundary conditions, so it is
interesting to ask what happens for other boundary conditions such as
free. Equation\ \eqref{BNPY} is actually rather surprising since it predicts that
finite-size corrections appear not when $\xi \sim L$, so $|T-T_c| \sim 1/ L^2$,
as one would expect, but
only when $\xi \sim L^{d/4}$, a larger scale, so $|T-T_c| \sim 1/L^{d/2}$, closer to
$T_c$ than expected. However,
as noted by Jones and Young~\cite{jones:05}, surely \textit{something} must
happen when $\xi \sim L$ with free boundary conditions, but what? In fact, in
an under-appreciated paper, Rudnick et al.~\cite{rudnick:85}, had previously argued
\textit{analytically} that that a temperature \textit{shift} of order $1/L^2$
has to be included with free boundary conditions, in addition to a rounding of
order $1/L^{d/2}$.

Even in the early days of
FSS~\cite{fisher:71,fisher:72b}, the possibility that a ``shift'' exponent could
be different from the ``rounding'' exponent was allowed for. To explain what
this means, note that the exponents $2$ in Eq.~\eqref{naive} and $d/2$ in
Eq.~\eqref{BNPY} are ``rounding'' exponents since they control the range of
temperature over which a singularity is rounded out ($L^{-2}$ and $L^{-d/2}$
respectively). To define the ``shift'' exponent we first define, for each size, a
``finite-size pseudocritical temperature''
$T_L$ by, for example, the location of the peak in some
susceptibility, or the temperature where the Binder ratio has a specified
value. The difference $T_c - T_L$ goes to zero for $L  \to \infty$ like
\begin{equation}
T_c - T_L = {A \over L^\lambda} \, ,
\label{TL}
\end{equation}
which is the desired definition of the shift exponent $\lambda$. The
precise value of $T_L$ depends on which criterion is used to define it, but
the exponent $\lambda$ is expected to be independent of the definition.
Whether or not the amplitude $A$ depends on the quantity used to define the shift
will be discussed in Sec.~\ref{sec:free}.
If $\lambda$ is less than
the rounding exponent, which will turn out to be the case for free boundary
conditions, then the shift is \textit{larger} than the rounding, so
we need to
modify Eq.~\eqref{BNPY} to
\begin{subequations}
\begin{align}
\chi(L, T) &= L^{y_T^\star} \, \overline{X}\left(L^{y_T^\star}\, (T-T_L)\right),
\label{chi_withshift}
\\
g(L, T) &= \overline{g}\left(L^{y_T^\star}\, (T-T_L)\right) \quad \text{(free, {\bf k}=0)} ,
\label{g_withshift}
\end{align}
\label{withshift}
\end{subequations}
in which the argument of the scaling function
involves the difference between $T$ and the ``finite-size pseudocritical
temperature'' $T_L$, and $y_T^\star = d/2$, see Eq.~\eqref{yTstar}.
The criterion that the shift is given by 
the condition $\xi \sim
L$ yields $\lambda = 2$, as proposed by Rudnick et al.\ \cite{rudnick:85} and
confirmed in simulations by Berche et al.\ \cite{berche:12}.
As with Eq.~\eqref{chi_standard}, we must have $\overline{X}(x) \propto
x^{-1}$ for $x \to \infty$ in order
to recover the bulk behavior above $T_c$. If we set $T = T_c$ then
$L^{d/2}\, (T-T_L) = A L^{d/2-2}$ which is large so we can use this limiting
behavior to get
\begin{equation}
\chi(L, T_c) \propto L^2 
\quad \text{(free, {\bf k}=0)} \, ,
\label{chiTc}
\end{equation}
a result which has been shown rigorously~\cite{watson:73}. Hence, in contrast to
Berche et al.~\cite{berche:12}, we propose that the region at the bulk $T_c$ 
\textit{is} part of the scaling function. Similarly, for the Binder ratio,
$\overline{g}(x) \propto 1/x^2$ for $x \to \infty$, which gives
\begin{equation}
g(L, T_c) \propto {1 \over L^{d-4}} 
\quad \text{(free, {\bf k}=0)} \, .
\label{gTc}
\end{equation}
With periodic boundary conditions, the intersection of the data for $g$
provides a convenient estimate of $T_c$, but, as Eq.~\eqref{gTc} shows,
this method cannot be used
for free boundary conditions
because $g$ vanishes at $T_c$ for $L \to \infty$. In fact, we shall see from
the numerical data in Sec.~\ref{sec:free} that
there are no intersections at all. However, we will not be able to verify
the precise form in Eq.~\eqref{gTc} because the values for $g$ at $T_c$ are so
small that the signal is lost in the noise.

So far we have discussed only ${\bf k} =0$ fluctuations. However, it is also
necessary to discuss fluctuations at ${\bf k} \ne 0$, since we need these to
determine the spatial decay of the correlation functions. Of particular
importance is the decay of the correlations at $T_c$, which fall off with
distance like $1/r^{d-2 + \eta}$, where the mean field value of
the exponent $\eta$ is
zero. In the mean field regime, the fluctuations of the ${\bf k} \ne 0$
modes are Gaussian so the Binder ratio is always zero. 
For the wave-vector dependent susceptibility we shall argue
that standard FSS,
Eq.~\eqref{naive}, holds for both boundary conditions, i.e.
\begin{equation}
\chi(k, L, T) = L^2 \widetilde{X}(L^2 \, (T - T_c), kL),
\ \text{(both bc's, ${\bf k} \ne 0$)}  ,
\label{kne0}
\end{equation}
where we have put the explicit $k$ dependence in a natural way as a second
argument of the scaling function. For free boundary conditions, the
Fourier modes are not plane waves,
see Sec.~\ref{sec:model}, and, by ${\bf k } \ne 0$, we really mean
modes that are orthogonal to the uniform magnetization and so do not develop a
non-zero expectation value below $T_c$. 

If we fix $T = T_c$ in
Eq.~\eqref{kne0} and consider $k L \gg 1$ then the size dependence 
must drop out
so $\widetilde{X}(0, y) \propto y^{-2}$ and hence
\begin{equation}
\chi(k, L, T_c) \propto k^{-2} \quad (k L \gg 1)\, .
\label{chikTc}
\end{equation}
Consequently, in real space, correlations fall of as $r^{-(d-2)}$, i.e.\ $\eta = 0$.
It follows that non-standard FSS only affects
the ${\bf k} =0$ mode and just
gives a larger baseline, $\sim 1/L^{d/2}$ rather than $1/L^{d-2}$, above
which the power law decay sits.  We therefore do not see the need for the second
$\eta$-like exponent proposed in Ref.~\cite{kenna:14}.

While Eq.~\eqref{kne0} does not seem to have been stated in the literature
before, to our knowledge, it is actually quite natural. The dangerous
irrelevant variable, which is the quartic coupling in the Ginzburg Landau
Wilson effective Hamiltonian, is needed to control the expectation value of
the (${\bf k}=0$) order parameter, which leads to non-standard FSS for ${\bf
k} = 0$ fluctuations. However, ${\bf k} \ne 0$ fluctuations (more precisely,
fluctuations which do not acquire a non-zero expectation value) are not
affected by the dangerous irrelevant variable, and consequently have standard
FSS.

The plan of this paper is as follows. In Sec.~\ref{sec:model} we define the
model to be simulated and the quantities we calculate. To incorporate
\textit{corrections} to FSS we use the quotient method which is described in
Sec.~\ref{sec:quot}. The numerical results for periodic boundary conditions
are presented in Sec.~\ref{sec:per} while those for free boundary conditions
are in Sec.~\ref{sec:free}.  We briefly summarize our conclusions in
Sec.~\ref{sec:conclusions}.

\section{Model}
\label{sec:model}
We consider an Ising model in $d=5$ dimensions with Hamiltonian
\begin{equation}
\mathcal{H} = -\sum_{\langle i , j \rangle} J_{i j} S_ i S_ j \, ,
\end{equation}
where $J_{i, j} = 1$ if $i$ and $j$ are nearest neighbors and zero otherwise,
and the spins $S_i$ take values $\pm 1$. The number of spins is $N = L^5$ and
we perform simulations with periodic and free boundary conditions.
Previous simulations have determined the transition temperature very
precisely,
finding~\cite{luijten:99}
\begin{equation}
T_c = 8.77846(3)\, . 
\label{Tc}
\end{equation}
We simulate this model very efficiently using the Wolff~\cite{wolff:89}
cluster algorithm, with which we can study sizes up to $L = 36$ (which
has around $60$ million spins). 

We calculate various moments of the uniform magnetization per spin
\begin{equation}
m = {1 \over L^d} \sum_{i=1}^N S_i\, ,
\label{uniform_m}
\end{equation}
as well as the uniform susceptibility~\cite{subtraction}
\begin{equation}
\chi = L^d \langle m^2 \rangle\, ,
\label{chi}
\end{equation}
and the Binder ratio
\begin{equation}
g = {1 \over 2} \, \left( 3 - {\langle m^4 \rangle \over \langle m^2
\rangle^2}
\right) \, .
\label{binder}
\end{equation}

In addition we compute the Fourier transformed susceptibilities
\begin{equation}
\chi({\bf k}) = L^d \langle |m({\bf k})|^2 \rangle \, ,
\end{equation}
in which the Fourier transformed magnetization, $m({\bf k})$, is defined
differently for periodic and free boundary conditions as follows.

For periodic boundary conditions the Fourier modes are plane waves so we have 
\begin{equation}
m({\bf k}) = {1 \over N} \sum_i e^{i {\bf k} \cdot {\bf r}}
\, S_i ,  \ \ \text{(periodic)},
\end{equation}
\label{mk_periodic}
where
\begin{equation}
k_\alpha = {2 \pi n_\alpha / L}, \ \ \text{(periodic)},
\label{kalpha_periodic}
\end{equation}
with $n_\alpha = 0, 1, \cdots, L-1$ and
$\alpha$ denotes a Cartesian coordinate. 

For free boundary conditions, the Fourier modes are sine waves,
\begin{equation}
m({\bf k}) = {1 \over N} \sum_i
\Bigl[\, \prod_{\alpha = 1}^d 
\sin\left(k_\alpha r_{i,\alpha}\right)\,\Bigr]
S_i ,  \ \ \text{(free)},
\label{mk_free}
\end{equation}
where
\begin{equation}
k_\alpha = {\pi n_\alpha /(L + 1)}, \ \ \text{(free)},
\label{kalpha_free}
\end{equation}
with $n_\alpha = 1, 2, \cdots, L$ and the components of the lattice
position, $r_{i,\alpha}$, also run over values
$1,2, \cdots, L$.
There is zero contribution to the sum 
in Eq.~\eqref{mk_free} if we set 
$r_{i,\alpha} = 0$ or $L+1$, so Eqs.~\eqref{mk_free} and\eqref{kalpha_free}
correctly incorporate free boundary conditions.

Note that ${\bf k} = 0$ is not
an allowed wavevector with free boundary conditions
so the uniform magnetization in Eq.~\eqref{uniform_m} does not
correspond to a single Fourier mode.
Note, too, that wavevectors with all $n_\alpha$ odd, have a projection on to
the uniform magnetization and so will acquire a non-zero expectation value
below $T_c$ in the thermodynamic limit.  They will therefore be subject to the
non-standard FSS in Eq.~\eqref{BNPY}. However, if any of the $n_\alpha$ are
even, there is no projection onto the uniform magnetization, so they will not
acquire an expectation value below $T_c$ and will therefore be subject to the
standard FSS in Eq.~\eqref{kne0}.

\section{The quotient method}
\label{sec:quot}

The discussion in Sec.~\ref{sec:intro} assumed that the sizes are
sufficiently large and $T$ sufficiently close to $T_c$ that the given FSS
formulae fit the data to high accuracy. For free boundary conditions, however,
\textit{corrections} to FSS are quite large and we need to include them in the
analysis. In this section we describe the method we used to include the \textit{leading}
correction to FSS.  

\begin{figure*}[!tb]
\begin{center}
\includegraphics[width=5.7cm]{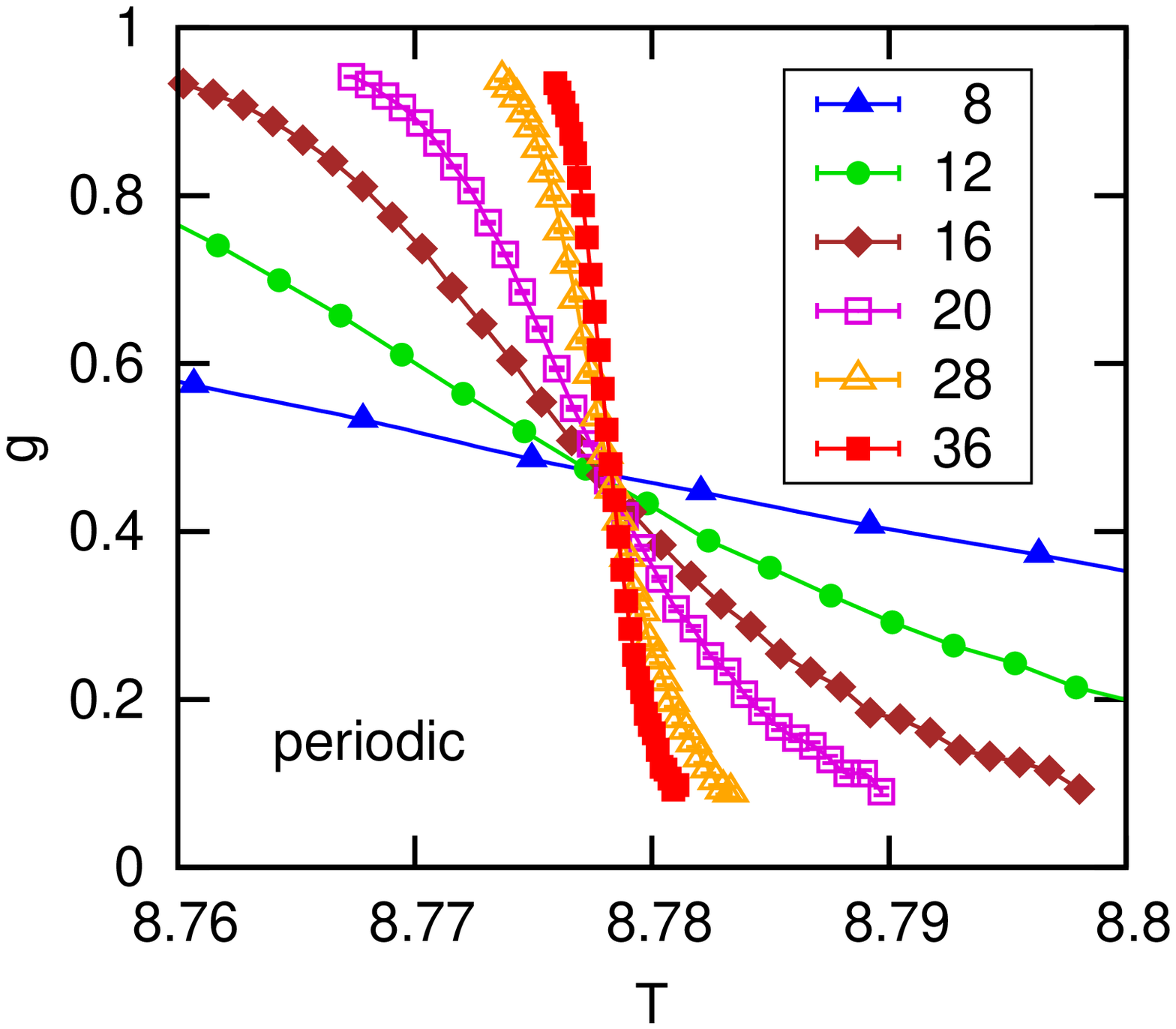}
\includegraphics[width=5.7cm]{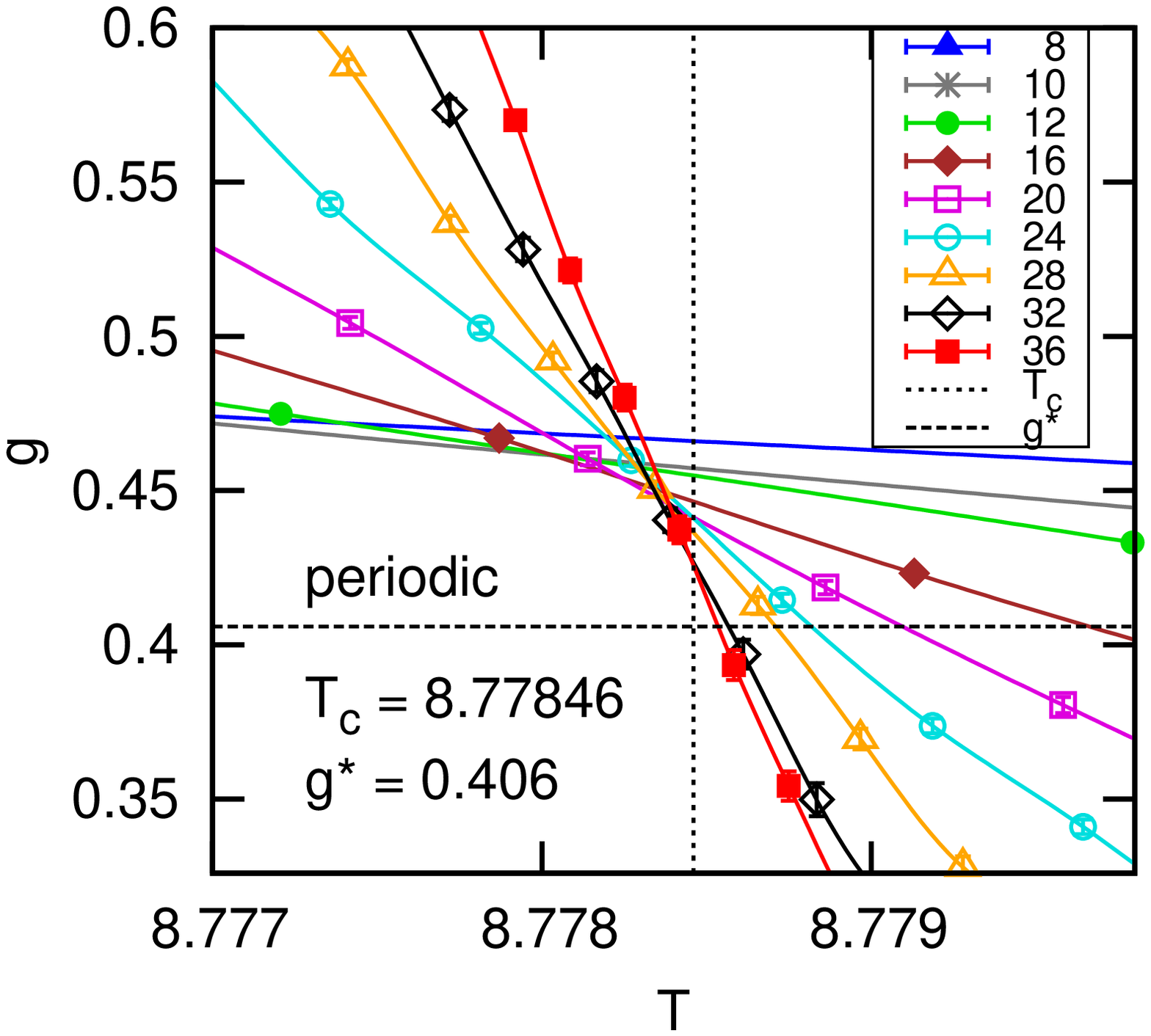}
\includegraphics[width=5.7cm]{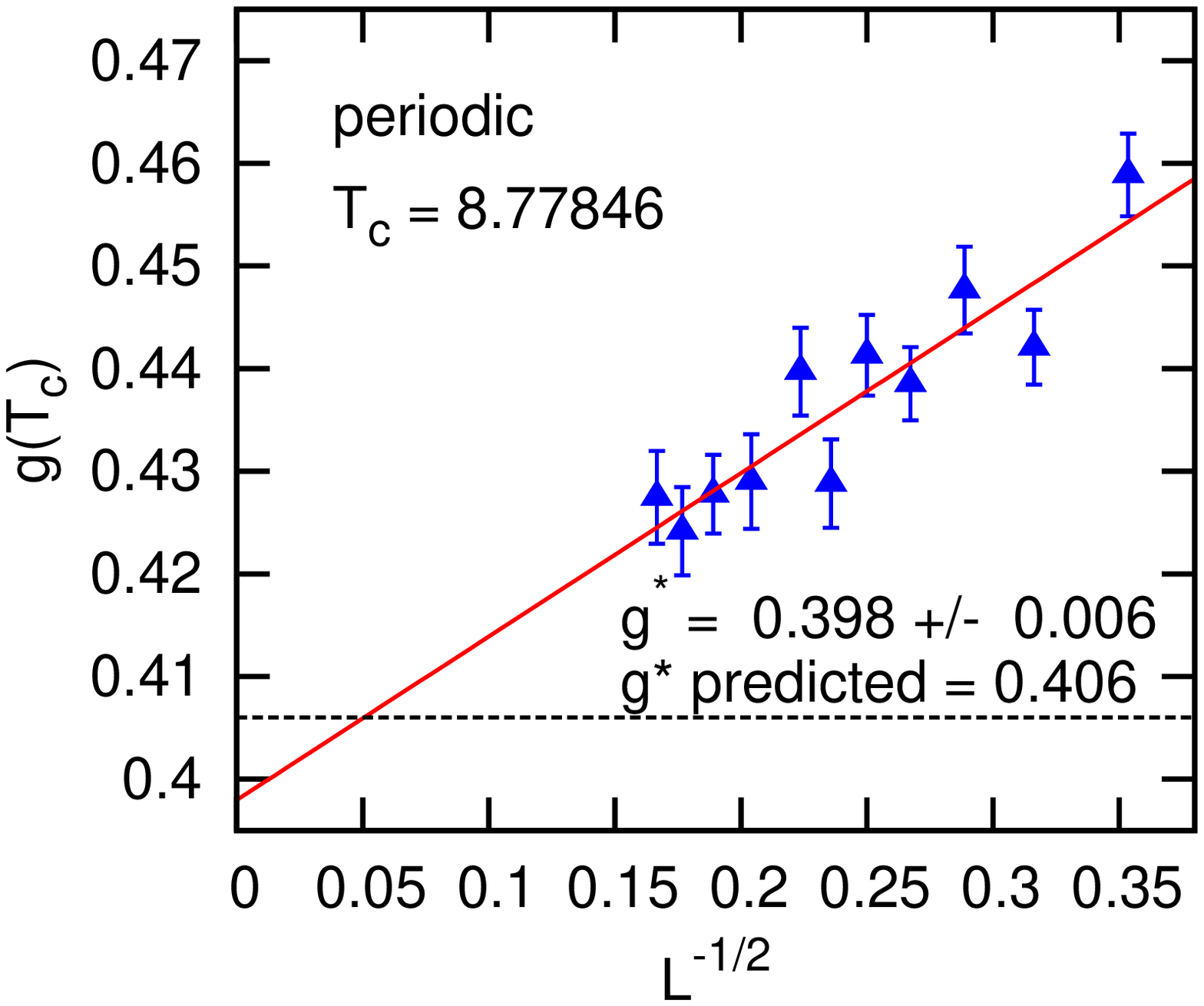}
\caption{(Color online) 
The left panel shows an overview of our results for the Binder ratio $g$ for
periodic boundary conditions. In the middle panel, which shows an expanded view
near $T_c$, the dashed vertical line indicates $T_c$
given by Eq.~\eqref{Tc}, and the dashed horizontal line indicates the
universal value for
$g_c$ given by Eq.~\eqref{gc}. The right panel shows an additional set of
results for $g$ taken precisely at $T_c$, plotted against $L^{-\omega'}$ with
$\omega' = 1/2$, see Eq.~\eqref{omega'}, and a straight-line fit
indicating an extrapolated value for $L \to \infty$
consistent with ($1.3 \sigma$
difference) the exact result. The quality of fit factor~\cite{press:92} is
$Q=0.252$.
\label{fig:k0_per}}
\end{center}
\end{figure*}

A convenient way to extract the leading scaling behavior from the data, 
in the presence of corrections, is the
quotient method~\cite{ballesteros:96}, which is a more modern version of
Nightingale's~\cite{nightingale:76}
phenomenological scaling.
As an example, consider the deviation of the pseudocritical
temperature $T_L$ from $T_c$ for which the FSS expression is given in
Eq.~\eqref{TL}. Including
the leading correction to scaling, which involves a universal exponent
$\omega$, one has
\begin{equation}
\Delta T(L) \equiv T_c - T_L  = {A \over L^\lambda} \, \left(1 + {B \over
L^\omega} \right) \, .
\label{correction}
\end{equation}
We determine the quotient $Q[\Delta T]$ by taking the log of the
ratio of the result for
sizes $L$ and $ s L$, where $s$ is a simple rational fraction like $2$ or $3/2$,
and divide by $\ln s$,
i.e.
\begin{equation}
Q_{s,L}[\Delta T] = {1\over \ln s } \, \ln \left({\Delta T(s L) \over \Delta T(L)}
\right) \, .
\label{quot}
\end{equation}
According to Eq.~\eqref{correction} we have, for large $L$,
\begin{equation}
Q_{s,L}[\Delta T] = -\lambda + {C_s \over L^\omega} \, ,
\label{Q_fit}
\end{equation}
where 
\begin{equation}
C_s = {s^{-\omega} - 1 \over \ln s} \, B \, .
\end{equation}
If the data is of sufficient quality, we can fit all the unknown parameters.
In Eq.~\eqref{Q_fit} these would be
$\lambda, \omega$ and $C_s$. In most cases, however, we
will need to assume the predicted value for the correction exponent $\omega$,
see below, and just fit to the other parameters.

According to the renormalization group, for $d > d_u = 4$, the leading
irrelevant variable has scaling dimension
\begin{equation}
\omega = d - 4 \, .
\label{omega}
\end{equation}
However, for ${\bf k} =0$ fluctuations and periodic boundary conditions, it
was shown in Ref.~\cite{brezin:85} that there is an additional, and larger,
correction for finite-size effects, with an exponent given by
\begin{equation}
\omega' = {d - 4 \over 2} \, .
\label{omega'}
\end{equation}
An intuitive way to see this is to note that the ``naive'' variation of
of $\chi$ with $L$ at the
critical point, $\chi \propto L^2$ see Eq.~\eqref{chi_naive}, although not the
dominant contribution (which is $L^{d/2}$, as shown in Eq.~\eqref{chi_BNPY}), is
nonetheless still
present as a correction. This correction
is down by a factor of $L^{2-d/2} \ (=L^{-\omega'})$ relative to the dominant
term. We
shall therefore use $\omega'$ rather than $\omega$ in considering corrections
to scaling for susceptibilities which scale with $L$ to the power $d/2$ rather than
$2$.

For some of our data we will also need subleading corrections to FSS for which
there are several contributions. One of these is the square of the leading
contribution. To avoid having too many fit parameters, this is the form we
shall assume, i.e.  when we include subleading corrections to scaling we will
do a parabolic fit in $1/L^\omega$ (or $1/L^{\omega'}$ as the case may be). 

A subtlety arises in doing fits to data for quotients, for example to
determine the parameters $\lambda, \omega$ and $C_s$ in Eq.~\eqref{Q_fit}. The
reason is that the same set of simulational data
may be used in the determination of more than one data point in the fit. 
For example, with $s=2$ the $L=16$ simulation data is incorporated into 
the pairs $(8, 16)$ \textit{and} $(16, 32)$.
Furthermore, we will do combined fits
incorporating data for
two different values of $s$ ($s=2$ and $3/2$), using the \textit{same}
exponents (since they are universal), but with 
different amplitudes (because they are not universal). This has the advantage of
increasing the number of data points in the fit by more than the number of parameters.
Again the same set of
simulational data is used to determine different
data points in the fit. Hence the
different quotient values being fitted are \textit{not statistically
independent.}  The best estimate of the fitting parameters should include
these correlations~\cite{ballesteros:96,ballesteros:98,weigel:09}.
In other words,
if a data point is $(x_i, y_i)$, and the fitting function is $u(x)$, which
depends on certain fitting parameters, those parameters should be determined by
minimizing
\begin{equation}
\chi^2 = \sum_{i, j} [y_i - u(x_i)]\, \left(C^{-1}\right)_{ij} \, 
[y_j - u(x_j)]\, , 
\end{equation}
where
\begin{equation}
C_{ij} = \langle y_i \, y_j \rangle -  \langle y_i \rangle \, \langle y_j
\rangle \, ,
\end{equation}
is the covariance matrix of the data.
We determine the elements of the covariance matrix
by a bootstrap analysis~\cite{newman:99,young:12}.
If there  are substantial correlations in many
elements, the covariance matrix can become singular, and where this happened we
projected on to the eigenvectors of the covariance matrix whose eigenvalues
are not (close to) zero, ignoring eigenvectors corresponding to zero
eigenvalues. The effective number of independent data points is then the rank
of the covariance matrix (the number of non-zero eigenvalues).

\begin{figure}[!tb]
\begin{center}
\includegraphics[width=\columnwidth]{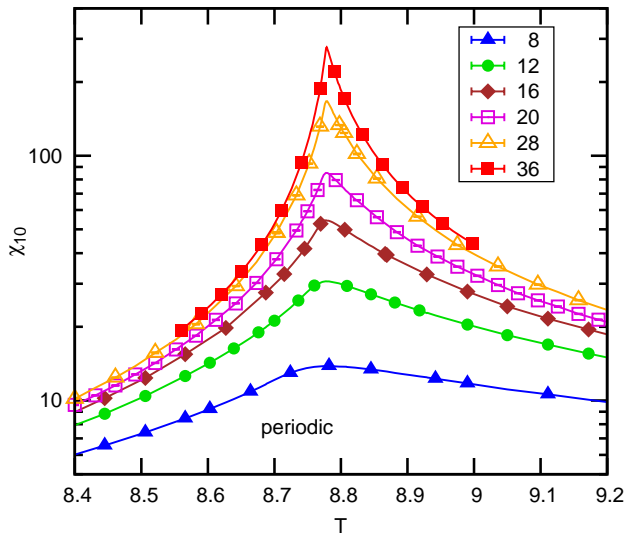}
\caption{(Color online) 
Susceptibility of $\chi({\bf k})$ for ${\bf k} L/(2\pi) = (1,0,0,0,0)$, which
we abbreviate to $\chi_{21}$,
for periodic boundary conditions. For clarity, only a representative selection of data
points is shown but the lines go through all the points.
\label{fig:chi_10_5d_per}
}
\end{center}
\end{figure}

\section{Results: periodic boundary conditions}
\label{sec:per}

\subsection{$\boldsymbol{k = 0}$ fluctuations}
\label{sec:per_k0}

We shall be brief here, since there is no dispute that the FSS scaling in
Eq.~\eqref{BNPY} is correct, but will show some results for completeness.

The left hand panel of Fig.~\ref{fig:k0_per} presents an overview of our data
for the Binder ratio $g$, showing intersections at, or close to, the
transition temperature $T_c$ given in Eq.~\eqref{Tc}. The expanded view in the
middle panel shows that the intersections for different pairs of sizes do not
occur at exactly the same, indicating corrections to scaling. In fact, the
data for smaller sizes have an approximate intersection at a value larger than
the exact, universal value of ~\cite{brezin:85}
\begin{equation}
g_c = {1\over 2}\left(3 - {\Gamma^4({1 \over 4}) \over 8\pi^2} \right) =
0.40578.
\label{gc}
\end{equation}
However, for larger sizes the intersections occur at smaller values of $g$.
The right hand panel of Fig.~\ref{fig:k0_per} shows an additional set of data
taken at precisely $T = T_c$, plotted against $L^{-\omega'}$ with the
correction exponent given by $\omega' = 1/2$, see the discussion in
Sec.~\ref{sec:quot}.
The data decreases to a value consistent with Eq.~\eqref{gc} for $L
\to\infty$.  As noted by
other authors, the effect of a fairly slow correction to scaling exponent,
$\omega = 1/2$, combined, evidently, with a fairly large correction amplitude,
has made it very difficult to obtain the known exact result for $g_c$
from numerics. This
should serve as a cautionary tale when applying FSS to other problems where
the exact answer is not known.

\subsection{$\boldsymbol{k \ne 0}$ fluctuations}
\label{sec:per_kn0}
The data for $\chi({\bf k})$ for ${\bf k} L/(2\pi) = (1,0,0,0,0)$ is shown in
Fig.~\ref{fig:chi_10_5d_per}.
Note that
the Fourier components at non-zero wavevector do not
develop order below $T_c$, and so what we define as $\chi({\bf k})$ really is
the susceptibility below $T_c$ as well as above it (unlike the ${\bf k=0}$
susceptibility~\cite{subtraction}), and consequently the data has a peak, 
whereas the uniform ``susceptibility''
plotted in Fig.~\ref{fig:chi_free} below (for free boundary conditions), 
continues to increase below $T_c$.

\begin{figure}[!tb]
\begin{center}
\includegraphics[width=\columnwidth]{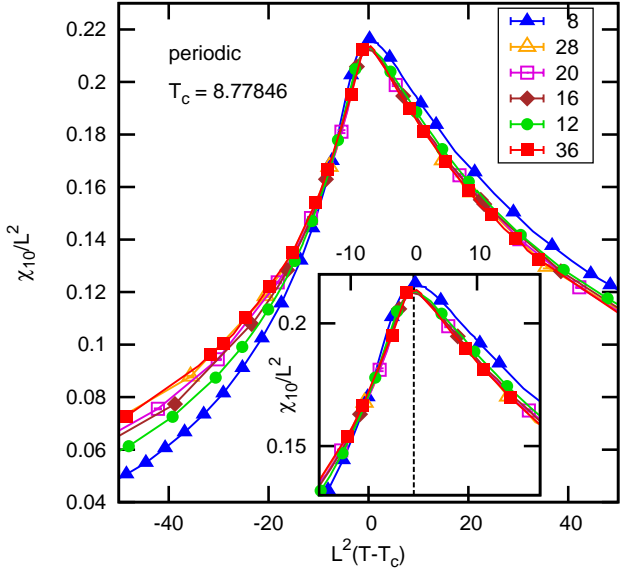}
\caption{(Color online) 
A scaling plot of
the susceptibility of the data in Fig.~\ref{fig:chi_10_5d_per}.
The inset shows an enlarged view near $T_c$. The
horizontal axis is $L^2(T-T_c)$ for both plots. 
\label{fig:chi_10_scale_per}}
\end{center}
\end{figure}

\begin{figure}[!tb]
\begin{center}
\includegraphics[width=\columnwidth]{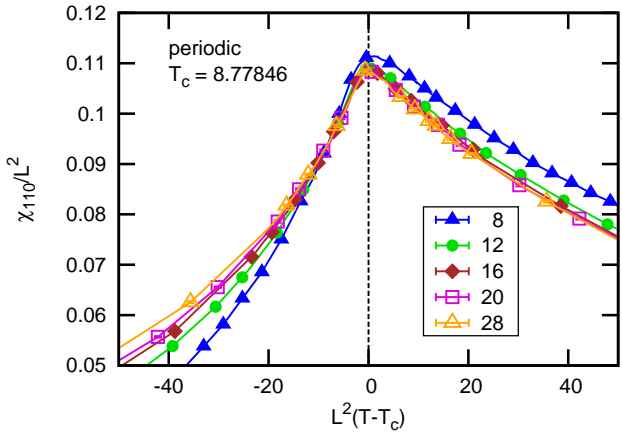}
\caption{(Color online) 
A scaling plot of
the susceptibility $\chi({\bf k})$ for ${\bf k}L/(2\pi) = (1,1,0,0,0)$,
which we abbreviate to $\chi_{110}$, for periodic
boundary conditions. 
\label{fig:chi_110_scale_per}}
\end{center}
\end{figure}

A scaling plot of the data is shown in Fig.~\ref{fig:chi_10_scale_per}
according to the standard FSS in Eq.~\eqref{kne0}. Apart from the smallest size,
$L=8$, near $T_c$
the data scales very well. Going further away from $T_c$ on the low-$T$ side, we
see bigger corrections. However, this is unsurprising since FSS is only
supposed to work for $T$ close to $T_c$.

If go to larger $k$-values we get a similar picture but with bigger
corrections to scaling, as shown in Fig.~\ref{fig:chi_110_scale_per} for
${\bf k} L/(2 \pi) = (1,1,0,0,0)$.
It is expected that corrections to scaling become
\textit{relatively} bigger for larger $k$ because the signal is less divergent
in this case and so is more easily affected by corrections.

\begin{figure}[!tb]
\begin{center}
\includegraphics[width=\columnwidth]{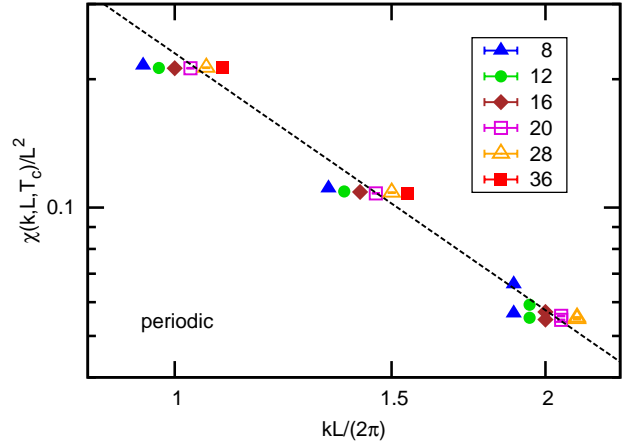}
\caption{(Color online) 
The values of $\chi({\bf k})/L^2$ at $T_c$ for periodic boundary conditions.
The points for different sizes and a single $k$ are displaced
slightly horizontally so they can be distinguished. Data is shown for three
different values of the $x$-axis: $1, \sqrt{2}$ and $2$. There are actually two
different wavevectors for $k L / (2\pi) = 2$, namely those
with ${\bf k} L/(2\pi) = (2,0,0,0,0)$
and $(1,1,1,1,0)$. These two agree well except for
the smaller sizes, showing that the fluctuations are isotropic at long
wavelength. The dashed line has slope $-2$ indicating that the expected
$k^{-2}$ behavior in Eq.~\eqref{chikTc} sets in even for small values of $k L$.
\label{fig:chik_Tc_5d_per}}
\end{center}
\end{figure}

Figure \ref{fig:chik_Tc_5d_per} shows the behavior of $\chi({\bf k})/L^2$ at
$T_c$ showing that it is a function of the product $k L$ as expected, see
Eq.~\eqref{kne0}. The dashed line has slope $-2$ indicating that the expected
$k^{-2}$ behavior in Eq.~\eqref{chikTc} sets in even for small values of $k
L$.

\begin{figure}[!tb]
\begin{center}
\includegraphics[width=\columnwidth]{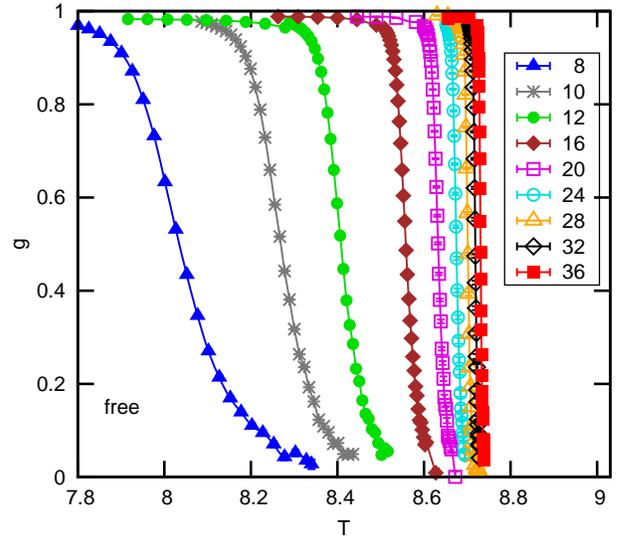}
\caption{(Color online) 
An overview of our results for the Binder ratio $g$ for
free boundary conditions.  Note that there is no sign of any intersections and
there is a large shift to lower temperatures for the smaller sizes.
\label{fig:g_k0_free}}
\end{center}
\end{figure}

\section{Results: free boundary conditions}
\label{sec:free}

Since corrections to scaling are larger for free boundary conditions than for
periodic boundary conditions, in this section we shall make extensive use of the quotient 
method described in Sec.~\ref{sec:quot} to incorporate the leading
correction.

\subsection{$\boldsymbol{k = 0}$ fluctuations}
\label{sec:free_k0}

An overview of our results for the Binder ratio is shown in
Fig.~\ref{fig:g_k0_free}. We do not find any intersections and the data is
shifted considerably to lower temperatures for smaller sizes. 

\begin{figure}[!tb]
\begin{center}
\includegraphics[width=\columnwidth]{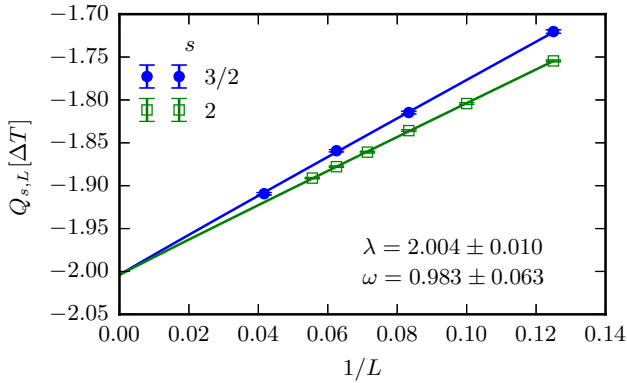}
\caption{(Color online) 
Quotients for $\Delta T(L)$, defined in Eq.~\eqref{correction},
used to determine the shift exponent $\lambda$ for free boundary
conditions. The data is fitted to Eq.~\eqref{Q_fit}, and the fitting
parameters are $\lambda, \omega$ (the same for both values of $s$) and
separate amplitudes $C_2$ and $C_{3/2}$ for the two $s$ values.
The quality of the linear fit is very good, $Q = 0.42$. 
\label{fig:shift_free}}
\end{center}
\end{figure}

In order to determine the shift exponent we define the 
pseudocritical temperature $T_L$ to be
where $g$ takes the value $1/2$, halfway between its limiting values of 0 and 1.
We subtract $T_c$ given in Eq.~\eqref{Tc} and determine the resulting
quotients for $\Delta T(L) \equiv T_c - T_L$ according to Eq.~\eqref{quot}. These
quotients are then fitted according to Eq.~\eqref{Q_fit}, as shown in
Fig.~\ref{fig:shift_free}. The quality of the data
is very good, the signal to noise is high, and we are able to fit all three
parameters $\lambda, \omega$ and the amplitude $C$.  The results for the
exponents are
\begin{equation}
\lambda = 2.004(10) , \qquad \omega = 0.98(6) \, .
\end{equation}
This value for the shift exponent is in precise agreement with the value 
$\lambda = 2$ proposed analytically in Ref.~\cite{rudnick:85}. There is also
excellent agreement between our value of the correction to scaling exponent
$\omega$ and the renormalization
group value of $1$.

\begin{figure}[!tb]
\begin{center}
\includegraphics[width=\columnwidth]{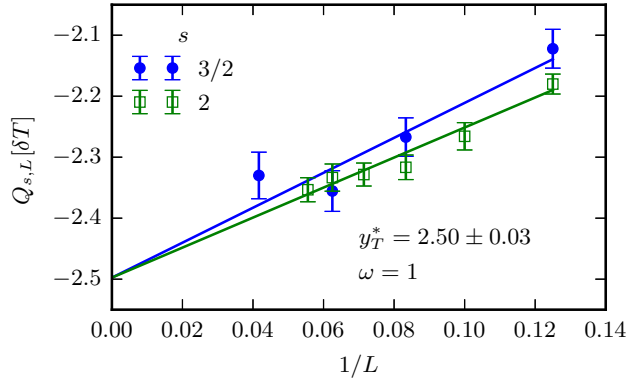}
\caption{(Color online) 
Quotients for $\delta T(L)$ defined in Eq.~\eqref{deltaT},
used to determine the width exponent $y_T^\star$ for free boundary
conditions. The data is fitted to Eq.~\eqref{QdT}, with fitting parameters
$y_T^\star, A_2$ and $A_{3/2}$. The correction exponent $\omega$ is fixed to
its expected value of 1. The quality of the straight-line fit is good; $Q = 0.50$.
\label{fig:width_free}
}
\end{center}
\end{figure}

We estimate the rounding by the range in temperature $\delta T(L)$ in which
$g$ varies between $0.25$ and $0.75$, i.e.
\begin{equation}
\delta T(L) = T(g=0.25) - T(g=0.75)\, .
\label{deltaT}
\end{equation}
Constructing the quotients and fitting to
\begin{equation}
Q_{s,L}[\delta T] = -y_T^\star + A_s / L^\omega
\label{QdT}
\end{equation}
we find that the data is insufficient to determine the three parameters, but
if we assume the RG value for the correction exponent, $\omega = 1$, then we
get a good fit which extrapolates to 
\begin{equation}
y_T^\star = 2.50(3)\, ,
\end{equation}
see Fig.~\ref{fig:width_free}, in precise agreement with the prediction $d/2$, 
see Eq.~\eqref{yTstar}.
Thus we have established the values of the
shift and rounding
exponents in Eq.~\eqref{g_withshift}.

\begin{figure}[!tb]
\begin{center}
\includegraphics[width=\columnwidth]{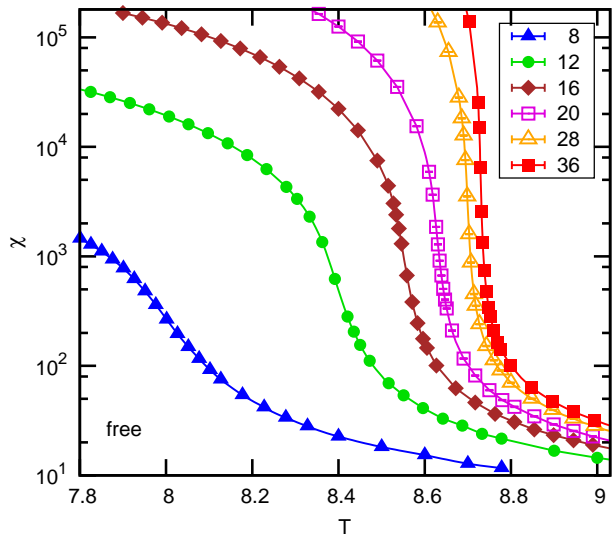}
\caption{(Color online) 
Data for $\chi$ for free boundary conditions.
Only a representative set of points are shown but
the lines go through all the points.
Note that with the definition in
Eq.~\eqref{chi} a term proportional to the square of the order parameter
is not subtracted off, so $\chi$ as defined is really only the
susceptibility above $T_c$, and continues to rise below $T_c$.
\label{fig:chi_free}
}
\end{center}
\end{figure}

\begin{figure}[!tb]
\begin{center}
\includegraphics[width=\columnwidth]{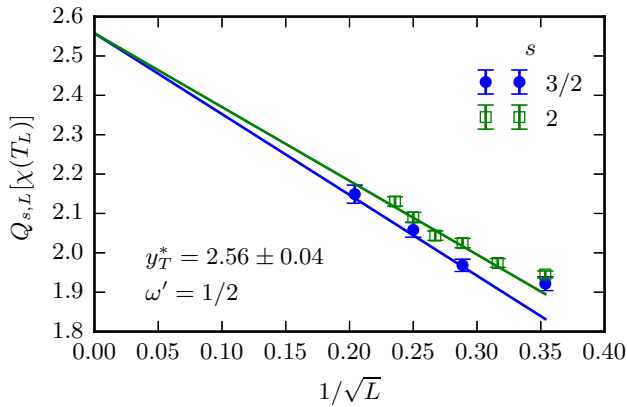}
\caption{(Color online) 
Quotients for value of $\chi$ at $T_L$ for free boundary conditions plotted
against $1/L^{\omega'}$ where the correction to scaling exponent 
$\omega'$ is fixed to the value $1/2$. According to Eq.~\eqref{chi_withshift} the
quotients should extrapolate to a value of $y_T^\star \ (=5/2)$ for $L \to
\infty$. 
The linear fit omits the right-hand point for each of
the data sets, and the three fitting parameters are
the value of $y_T$ and two amplitudes
of the correction, one for each value of $s$. The quality of the fit is good, $Q=0.30$.
\label{fig:chi_at_TL_free}
}
\end{center}
\end{figure}

What about the scaling of $\chi$ in Eq.~\eqref{chi_withshift}? The data for
$\chi$ is shown in Fig.~\ref{fig:chi_free}. 
We evaluated this at $T_L$, and did a quotient analysis which is shown
in Fig.~\ref{fig:chi_at_TL_free}. 
The data is insufficient to determine the correction to scaling exponent, so
we fixed it to the expected value $\omega' = 1/2$. The amplitude of the
correction term is large, but the data extrapolates to a value $2.56(4)$, very
close to the value of $y_T^\star = 5/2$
expected from to Eq.~\eqref{chi_withshift}.

\begin{figure}[!tb]
\begin{center}
\includegraphics[width=\columnwidth]{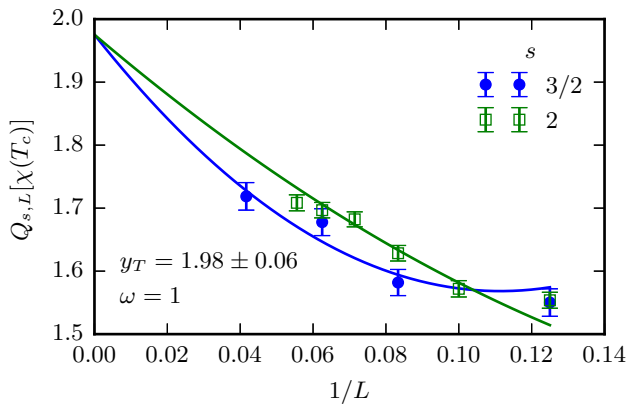}
\caption{(Color online) 
A quadratic fit for the quotients for value of $\chi$ at the bulk $T_c$ for
free boundary conditions against $1/L^\omega$ where the correction to
scaling exponent $\omega$ is fixed to the value $1$. According to Eq.~\eqref{chiTc},
the quotients should extrapolate to the value of $y_T\ (= 2)$. There are five
fitting parameters: $y_T$ and the amplitudes of the linear and quadratic
corrections for each $s$ value. The quality of the fit is
good, $Q = 0.43$.
\label{fig:chi0-Tc-quots-w1-quad}
}
\end{center}
\end{figure}

We can also evaluate $\chi$ at the bulk $T_c$. As shown in Eq.~\eqref{chiTc},
this is proportional to $L^2$, not $L^{d/2}$, and so, as discussed in
Sec.~\ref{sec:quot}.
we expect that the correction to scaling exponent will be $\omega \ (=1)$
rather than $\omega'\ (= 1/2)$.  Quotients of the results are plotted in
Fig.~\ref{fig:chi0-Tc-quots-w1-quad}. There are clearly subleading
corrections to scaling so we try a quadratic fit, with the result
$y_T = 1.97(6)$ in good agreement with the
expected value of $2$.
We note that corrections to scaling are quite large, which is not surprising
since
the values of $\chi$ at $T_c$ are quite small, and so
are more influenced by several corrections to scaling than the data at $T_L$
which $\chi$ is bigger. We also tried a linear fit omitting the smallest size
for each value of $s$ finding $1.89(2)$ with $Q = 0.30$ which differs by more
than the error bar from the value $2$. Nonetheless, the quadratic fit shows that,
although we have not determined
the exponent with which $\chi$ diverges at $T_c$ with great accuracy, it is,
at the very least, \textit{consistent} with the expected value in Eq.~\eqref{chiTc}.

\begin{figure}[!tb]
\begin{center}
\includegraphics[width=\columnwidth]{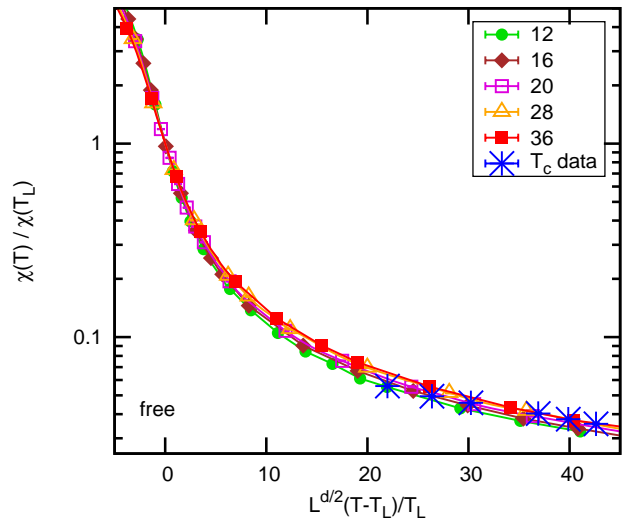}
\caption{(Color online) 
A scaling plot of the data for $\chi$ for free boundary conditions according
to Eq.~\eqref{chi_withshift}. Also shown is the data at $T_c$ which is seen to
lie on the
scaling function (within some small corrections.)
\label{fig:chi_5d_free_scale}
}
\end{center}
\end{figure}

Figure \ref{fig:chi_5d_free_scale} shows a scaling plot of $\chi(T)/\chi(T_L)$
against $L^{d/2} (T - T_L)/T_L$. We have seen in Fig.~\ref{fig:chi_at_TL_free}
that there are corrections to the expected $L^{d/2}$ behavior of $\chi$ at
$T_L$ for the range of sizes studied. Hence we divide  $\chi(T)$ by
$\chi(T_L)$ rather than by $L^{d/2}$ which appears in
Eq.~\eqref{chi_withshift}, to eliminate those corrections to scaling in
Fig.~\ref{fig:chi_5d_free_scale}.
According to Eq.~\eqref{chi_withshift} the data in
Fig.~\ref{fig:chi_5d_free_scale} should
collapse. There are some corrections to this, which
is not surprising since we are probing the scaling function over a big region,
but overall the data scales pretty well. Also shown are data at $T_c$, which appears
at different points for different sizes because $T_L$ is, of course, size
dependent. The larger the size, the  further to the right is the data point
for $T_c$.
This figure supports our claim that the data at $T_c$ \textit{is} included in
the scaling function in Eq.~\eqref{chi_withshift}.

\begin{figure}[!tb]
\begin{center}
\includegraphics[width=\columnwidth]{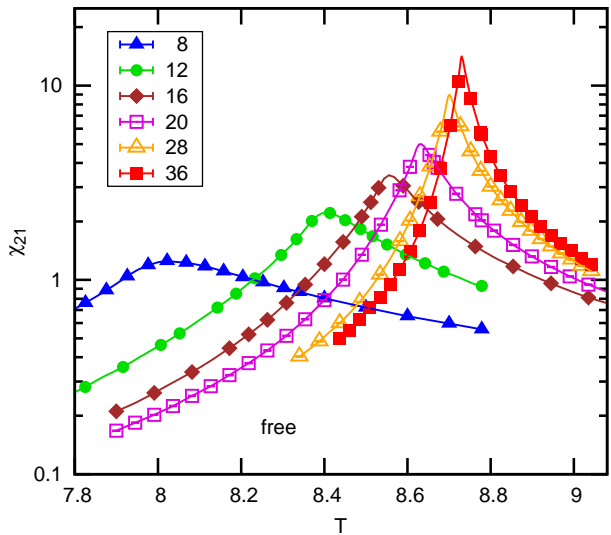}
\caption{(Color online) 
Data for $\chi({\bf k})$ for $(L+1){\bf k} / \pi =
(2,1,1,1,1)$ for 
free boundary conditions.
Only a representative set of points are shown but
the lines go through all the points.
\label{fig:chi_21_free}
}
\end{center}
\end{figure}

We have defined the pseudocritical temperatures $T_L$, and the resulting shift
exponent $\lambda$, from Eq.~\eqref{TL} by the temperature where the Binder ratio
takes the value $1/2$. Suppose we took a different criterion for  $T_L$, such
as the temperature where the Binder ratio has some other value, or where there
is a peak in some ${\bf k} \ne 0$ susceptibility such as that shown in
Fig.~\ref{fig:chi_21_free}. We note that the finite-size width varies as
$1/L^{d/2}$ so temperatures where the Binder ratio has a value between 0 and
1 would lie in this range, and so would only give a \textit{sub-leading}
contribution to the shift, the coefficient of $1/L^2$ remaining the
same. We expect that the same
shift amplitude would be obtained no matter what quantity is used to define
the shift for the following reason. Suppose we have a shift amplitude $A$
and pseudocritical temperatures $T_L$
determined from where
the Binder ratio is $1/2$ and a different amplitude $A'$, and correspondingly
different temperatures $T'_L$, determined by some
other criteria. Then the Binder ratio has scaling form in
Eq.~\eqref{g_withshift}, but if
we try to define it in terms of the alternative shift temperatures $T'_L$ we have
\begin{align}
g(L, T) &= \overline{g}\left(L^{d/2}\, (T-T_L)\right) \\
&= \overline{g}\left(L^{d/2}\, (T-T'_L) + (A' - A) L^{d/2-2}\, \right)\, .
\label{g_withshift'}
\end{align}
Hence, if different quantities give different shift amplitudes, the argument
of the scaling function would be shifted by an \textit{infinite} amount (for $L \to
\infty$) if we use the shift obtained from a different quantity.
This would be a clear violation of scaling. We postulate that this does not
happen and that there is a \textit{unique} shift amplitude for a given system.

Note, however, that we cannot rule out subleading corrections to the shift of
order $1/L^{d/2}$. As a result, the value of $g$ at $T_L$ according to
Eq.~\eqref{g_withshift} will depend on the
precise definition of $T_L$ and therefore \textit{not} be universal, unlike the
situation with periodic boundary conditions, see Eq.~\eqref{g_BNPY}. Hence one
can view
the replacement of Eqs.~\eqref{BNPY} by
Eqs.~\eqref{withshift} as a violation of standard finite-size scaling~\cite{rudnick:85}.
However, since the the
behavior of $\chi$, for example, is described by a single function both at
$T_c$ and $T_L$, we view Eqs.~\eqref{withshift} as representing a
\textit{modified FSS},
distinct from standard FSS in that it has different shift and scaling exponents.

\subsection{$\boldsymbol{k \ne 0}$ fluctuations}
\label{sec:free_kn0}

With free boundary conditions the Fourier modes are sine waves given by
Eq.~\eqref{kalpha_free}. Modes in which all the integers $n_\alpha$ are odd
have a projection on the uniform magnetization and so will acquire a non-zero
magnetization.  These will be therefore be affected by the dangerous
irrelevant variable and so have the same scaling as fluctuations of the
uniform magnetization, given in Eq.~\eqref{chi_withshift}. We therefore take the
smallest wavevector with an even $n_\alpha$, namely ${\bf n} = (2,1,1,1,1)$,
since this will not acquire a non-zero magnetization so we expect it to be
governed by the FSS in Eq.~\eqref{kne0}, i.e.\ with exponent $2$ rather than
$d/2$ which appears in Eq.~\eqref{chi_withshift}. We show the data in
Fig.~\ref{fig:chi_21_free}.

\begin{figure}[!tb]
\begin{center}
\includegraphics[width=\columnwidth]{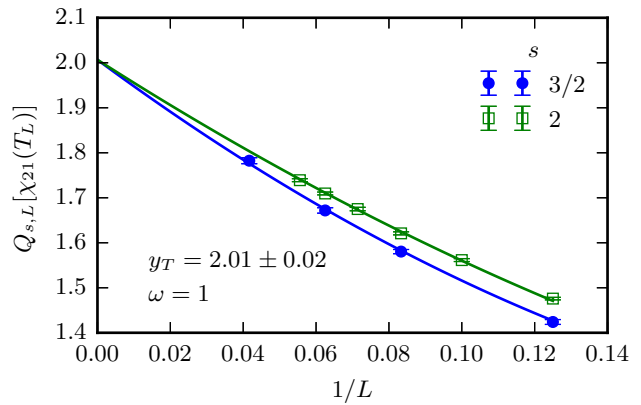}
\caption{(Color online) 
Quotients for the value of $\chi({\bf k})$ for $(L+1){\bf k} / \pi =
(2,1,1,1,1)$ at $T_L$ for free boundary conditions. The correction to scaling exponent
of $\omega = 1$ is taken.  According to Eq.~\eqref{kne0} the quotients should tend to the value
$y_T \ (=2)$ for $L \to \infty$. As in other quotient fits, we use the same values for the
exponents $y_T$ and $\omega$ for the two values of $s$, but different
amplitudes for the corrections to scaling. Here we use a quadratic fit which worked well, 
$Q = 0.53$. A linear fit gave a
an extrapolated value of $1.950(2)$ but with a poor quality of fit factor
$Q=0.002$.
\label{fig:chi21-Ts-quots-w1-quad}
}
\end{center}
\end{figure}

\begin{figure}[!tb]
\begin{center}
\includegraphics[width=\columnwidth]{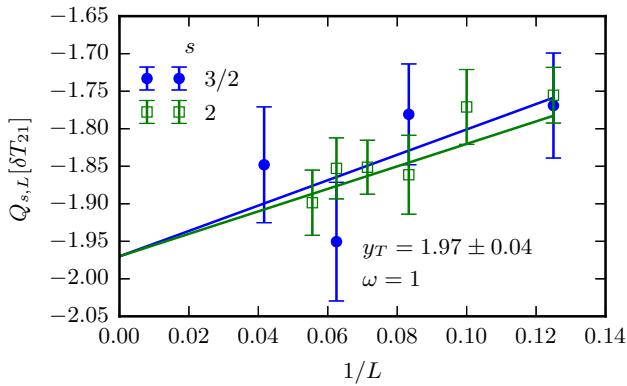}
\caption{(Color online) 
Quotients for the width of the peak in $\chi({\bf k})$ for $(L+1){\bf k} / \pi =
(2,1,1,1,1)$ for free boundary conditions. The correction to scaling exponent
of $\omega = 1$ is taken. According to Eq.~\eqref{kne0} the quotients should
tend to $-y_T \ (=-2)$ for $L \to\infty$. 
The amplitude of the correction to scaling is seen 
to be quite small in this case,
and the quality of linear fit is excellent: $Q = 0.67$.
\label{fig:chi21-width-quots-w1}
}
\end{center}
\end{figure}

According to Eq.~\eqref{kne0} the height of the peaks in
Fig.~\ref{fig:chi_21_free} should
scale as $L^2$ and the width should scale as $L^{-2}$. We define the width to
be the difference between the two temperatures where the susceptibility is $3/4$
of that at the maximum. The quotient analyses for height and width are shown in
Figs.~\ref{fig:chi21-Ts-quots-w1-quad} and
\ref{fig:chi21-width-quots-w1} respectively. For the height the (quadratic)
fit gives an extrapolated value of $2.010(24)$ which agrees with the expected
value of $y_T = 2$. As discussed in the caption to
Fig.~\ref{fig:chi21-Ts-quots-w1-quad} a linear fit gave a value
$1.950(2)$, close
to but slightly different from 2. However, the quality of fit factor
$Q= 0.002$~\cite{press:92} was unacceptably low, which is why we went to
a quadratic fit. For the data of the width in
Fig.~\ref{fig:chi21-width-quots-w1} the dependence on size is modest and we
find an extrapolated value of $-1.97(4)$ well consistent with the expected
value of $-y_T \ (=-2)$.

Consequently we have found strong evidence to support our claim that
Eq.~\eqref{kne0} applies to free boundary conditions. Note that since this FSS
scaling form uses $y_T \ (=2)$ and the deviation of $T_L$ from $T_c$ is
proportional to $1/L^2$, asymptotically 
we can use either $T_c$ or $T_L$ in Eq.~\eqref{kne0}.

\section{Summary and Conclusions}
\label{sec:conclusions}

Our main conclusions have already been discussed in the introduction so we will be
brief here. FSS above the upper critical dimension can be summarized by:
\begin{enumerate}
\item
The modified FSS form with exponents $d/2$ rather than $2$ only applies to
${\bf k=0}$ fluctuations. (For free boundaries, it applies to Fourier modes
which have a projection onto the uniform magnetization.) For all other
wavevectors, standard FSS with an exponent $2$ applies. As a result there is
only one exponent $\eta$ describing the power-law decay of correlations at
$T_c$ in contrast to recent claims.
\item
For free boundaries and at ${\bf k}=0$, the shift, with an exponent $2$, is
larger than the rounding, which has an exponent $d/2$. Using $T-T_L$, where
$T_L$ is the finite-size, pseudocritical temperature, rather than $T-T_c$,
as a scaling variable, the
data has a scaling form which incorporates both the behavior at $T_L$ where
$\chi \propto L^{d/2}$ and at the bulk $T_c$ where $\chi \propto L^2$.
\end{enumerate}

\begin{acknowledgments}
{
This work is supported in part by the National
Science Foundation under Grant
No.~DMR-1207036.  We also
acknowledge support from a Gutzwiller Fellowship at the Max Planck Institute
for the Physics of Complex Systems, Dresden.
}

\end{acknowledgments}

\bibliography{refs,comments}

\begin{thebibliography}{25}
\expandafter\ifx\csname natexlab\endcsname\relax\def\natexlab#1{#1}\fi
\expandafter\ifx\csname bibnamefont\endcsname\relax
  \def\bibnamefont#1{#1}\fi
\expandafter\ifx\csname bibfnamefont\endcsname\relax
  \def\bibfnamefont#1{#1}\fi
\expandafter\ifx\csname citenamefont\endcsname\relax
  \def\citenamefont#1{#1}\fi
\expandafter\ifx\csname url\endcsname\relax
  \def\url#1{\texttt{#1}}\fi
\expandafter\ifx\csname urlprefix\endcsname\relax\def\urlprefix{URL }\fi
\providecommand{\bibinfo}[2]{#2}
\providecommand{\eprint}[2][]{\url{#2}}

\bibitem[{\citenamefont{Fisher}(1971)}]{fisher:71}
\bibinfo{author}{\bibfnamefont{M.~E.} \bibnamefont{Fisher}},
  \emph{\bibinfo{title}{The theory of critical point singularities}}, in
  \emph{\bibinfo{booktitle}{Critical Phenomena, Proceedings of the 51st Enrico
  Fermi Summer School, Varenna}}, edited by
  \bibinfo{editor}{\bibfnamefont{M.~S.} \bibnamefont{Green}}
  (\bibinfo{publisher}{Academic Press}, \bibinfo{address}{New York},
  \bibinfo{year}{1971}), p.~\bibinfo{pages}{1}.

\bibitem[{\citenamefont{Fisher and Barber}(1972)}]{fisher:72b}
\bibinfo{author}{\bibfnamefont{M.~E.} \bibnamefont{Fisher}} \bibnamefont{and}
  \bibinfo{author}{\bibfnamefont{M.~N.} \bibnamefont{Barber}},
  \emph{\bibinfo{title}{Scaling theory for finite-size effects in the critical
  region}}, \bibinfo{journal}{Phys. Rev. Lett.} \textbf{\bibinfo{volume}{28}},
  \bibinfo{pages}{1516} (\bibinfo{year}{1972}).

\bibitem[{\citenamefont{Binder and Luijten}(2001)}]{binder:01}
\bibinfo{author}{\bibfnamefont{K.}~\bibnamefont{Binder}} \bibnamefont{and}
  \bibinfo{author}{\bibfnamefont{E.}~\bibnamefont{Luijten}},
  \emph{\bibinfo{title}{Monte {C}arlo tests of renormalization group
  predictions for critical phenomena in {I}sing models}},
  \bibinfo{journal}{Phys. Rep.} \textbf{\bibinfo{volume}{344}},
  \bibinfo{pages}{179} (\bibinfo{year}{2001}).

\bibitem[{pow()}]{power}
\bibinfo{note}{It is convenient to take $L/\xi$ to the power $1/\nu$ because
  then $T- T_c$ appears linearly, with the result that a single scaling
  function applies both above and below $T_c$.}

\bibitem[{\citenamefont{Binder}(1981)}]{binder:81}
\bibinfo{author}{\bibfnamefont{K.}~\bibnamefont{Binder}},
  \emph{\bibinfo{title}{Critical properties from {M}onte {C}arlo coarse
  graining and renormalization}}, \bibinfo{journal}{Phys. Rev. Lett.}
  \textbf{\bibinfo{volume}{47}}, \bibinfo{pages}{693} (\bibinfo{year}{1981}).

\bibitem[{\citenamefont{Binder et~al.}(1985)\citenamefont{Binder, Nauenberg,
  Privman, and Young}}]{binder:85}
\bibinfo{author}{\bibfnamefont{K.}~\bibnamefont{Binder}},
  \bibinfo{author}{\bibfnamefont{M.}~\bibnamefont{Nauenberg}},
  \bibinfo{author}{\bibfnamefont{V.}~\bibnamefont{Privman}}, \bibnamefont{and}
  \bibinfo{author}{\bibfnamefont{A.~P.} \bibnamefont{Young}},
  \emph{\bibinfo{title}{Finite-size tests of hyperscaling}},
  \bibinfo{journal}{Phys. Rev. B} \textbf{\bibinfo{volume}{31}},
  \bibinfo{pages}{1498} (\bibinfo{year}{1985}).

\bibitem[{\citenamefont{Br\'ezin and Zinn-Justin}(1985)}]{brezin:85}
\bibinfo{author}{\bibfnamefont{E.}~\bibnamefont{Br\'ezin}} \bibnamefont{and}
  \bibinfo{author}{\bibfnamefont{J.}~\bibnamefont{Zinn-Justin}},
  \emph{\bibinfo{title}{Finite size effects in phase transitions}},
  \bibinfo{journal}{Nucl. Phys. B} \textbf{\bibinfo{volume}{257}},
  \bibinfo{pages}{867} (\bibinfo{year}{1985}).

\bibitem[{\citenamefont{Luijten and Bl\"ote}(1996)}]{luijten:96}
\bibinfo{author}{\bibfnamefont{E.}~\bibnamefont{Luijten}} \bibnamefont{and}
  \bibinfo{author}{\bibfnamefont{H.~W.~J.} \bibnamefont{Bl\"ote}},
  \emph{\bibinfo{title}{Finite-size scaling and universality above the upper
  critical dimension}}, \bibinfo{journal}{Phys. Rev. Lett.}
  \textbf{\bibinfo{volume}{76}}, \bibinfo{pages}{1557} (\bibinfo{year}{1996}).

\bibitem[{\citenamefont{Parisi and Ruiz-Lorenzo}(1996)}]{parisi:96b}
\bibinfo{author}{\bibfnamefont{G.}~\bibnamefont{Parisi}} \bibnamefont{and}
  \bibinfo{author}{\bibfnamefont{J.~J.} \bibnamefont{Ruiz-Lorenzo}},
  \emph{\bibinfo{title}{Scaling above the upper critical dimension in {I}sing
  models}}, \bibinfo{journal}{Phys. Rev. B} \textbf{\bibinfo{volume}{54}},
  \bibinfo{pages}{R3698} (\bibinfo{year}{1996}).

\bibitem[{\citenamefont{Bl\"ote and Luijten}(1997)}]{blote:97}
\bibinfo{author}{\bibfnamefont{H.~W.~J.} \bibnamefont{Bl\"ote}}
  \bibnamefont{and} \bibinfo{author}{\bibfnamefont{E.}~\bibnamefont{Luijten}},
  \emph{\bibinfo{title}{Universality and the five-dimensional {I}sing model}},
  \bibinfo{journal}{EuroPhys. Lett.} \textbf{\bibinfo{volume}{38}},
  \bibinfo{pages}{565} (\bibinfo{year}{1997}).

\bibitem[{\citenamefont{Luijten et~al.}(1999)\citenamefont{Luijten, Binder, and
  Bl\"ote}}]{luijten:99}
\bibinfo{author}{\bibfnamefont{E.}~\bibnamefont{Luijten}},
  \bibinfo{author}{\bibfnamefont{K.}~\bibnamefont{Binder}}, \bibnamefont{and}
  \bibinfo{author}{\bibfnamefont{H.~W.~J.} \bibnamefont{Bl\"ote}},
  \emph{\bibinfo{title}{Finite-size scaling above the upper critical dimension
  revisited: the case of the five-dimensional {I}sing model}},
  \bibinfo{journal}{Eur. Phys. J. B} \textbf{\bibinfo{volume}{9}},
  \bibinfo{pages}{289} (\bibinfo{year}{1999}).

\bibitem[{\citenamefont{Jones and Young}(2005)}]{jones:05}
\bibinfo{author}{\bibfnamefont{J.~L.} \bibnamefont{Jones}} \bibnamefont{and}
  \bibinfo{author}{\bibfnamefont{A.~P.} \bibnamefont{Young}},
  \emph{\bibinfo{title}{Finite size scaling of the correlation length above the
  upper critical dimension}}, \bibinfo{journal}{Phys. Rev. B}
  \textbf{\bibinfo{volume}{71}}, \bibinfo{pages}{174438}
  (\bibinfo{year}{2005}), \eprint{(arXiv:cond-mat/0412150)}.

\bibitem[{\citenamefont{Rudnick et~al.}(1985)\citenamefont{Rudnick, Gaspari,
  and Privman}}]{rudnick:85}
\bibinfo{author}{\bibfnamefont{J.}~\bibnamefont{Rudnick}},
  \bibinfo{author}{\bibfnamefont{G.}~\bibnamefont{Gaspari}}, \bibnamefont{and}
  \bibinfo{author}{\bibfnamefont{V.}~\bibnamefont{Privman}},
  \emph{\bibinfo{title}{Effect of boundary conditions on the critical behavior
  of a finite high-dimensional {I}sing model}}, \bibinfo{journal}{Phys. Rev. B}
  \textbf{\bibinfo{volume}{32}}, \bibinfo{pages}{7594} (\bibinfo{year}{1985}).

\bibitem[{\citenamefont{Berche et~al.}(2012)\citenamefont{Berche, Kenna, and
  Walter}}]{berche:12}
\bibinfo{author}{\bibfnamefont{B.}~\bibnamefont{Berche}},
  \bibinfo{author}{\bibfnamefont{R.}~\bibnamefont{Kenna}}, \bibnamefont{and}
  \bibinfo{author}{\bibfnamefont{J.-C.} \bibnamefont{Walter}},
  \emph{\bibinfo{title}{Hyperscaling above the upper critical dimension}},
  \bibinfo{journal}{Nuclear Physics B} \textbf{\bibinfo{volume}{865}},
  \bibinfo{pages}{115} (\bibinfo{year}{2012}).

\bibitem[{\citenamefont{Watson}(1973)}]{watson:73}
\bibinfo{author}{\bibfnamefont{P.~G.} \bibnamefont{Watson}},
  \emph{\bibinfo{title}{Surface and size effects in lattice models}}, in
  \emph{\bibinfo{booktitle}{Phase Transitions and Critical Phenomena, Vol. 2}},
  edited by \bibinfo{editor}{\bibfnamefont{C.}~\bibnamefont{Domb}}
  \bibnamefont{and} \bibinfo{editor}{\bibfnamefont{M.}~\bibnamefont{Green}}
  (\bibinfo{publisher}{Academic Press}, \bibinfo{address}{London},
  \bibinfo{year}{1973}), p. \bibinfo{pages}{101}.

\bibitem[{\citenamefont{Kenna and Berche}(2014)}]{kenna:14}
\bibinfo{author}{\bibfnamefont{R.}~\bibnamefont{Kenna}} \bibnamefont{and}
  \bibinfo{author}{\bibfnamefont{B.}~\bibnamefont{Berche}},
  \emph{\bibinfo{title}{Fisher's scaling relation above the upper critical
  dimension}}, \bibinfo{journal}{Europhys. Lett.}
  \textbf{\bibinfo{volume}{105}}, \bibinfo{pages}{26005}
  (\bibinfo{year}{2014}).

\bibitem[{\citenamefont{Wolff}(1989)}]{wolff:89}
\bibinfo{author}{\bibfnamefont{U.}~\bibnamefont{Wolff}},
  \emph{\bibinfo{title}{Collective {M}onte {C}arlo updating for spin systems}},
  \bibinfo{journal}{Phys. Rev. Lett.} \textbf{\bibinfo{volume}{62}},
  \bibinfo{pages}{361} (\bibinfo{year}{1989}).

\bibitem[{sub()}]{subtraction}
\bibinfo{note}{This expression differs from the standard expression for the
  susceptibility $\chi= \beta L^d\left(\langle m^2\rangle - \langle m
  \rangle^2\right)$ in two ways. The first trivial difference is that we omit
  the factor of $\beta$ which is conventional in critical phenomena studies.
  Secondly, and not so trivially, we ignore the subtracted term, which is hard
  to compute reliably by Monte Carlo since one has to apply a field $h$ and
  take the limit $h \to 0$ \textit{after} the limit $L \to\infty$. Hence the
  quantity we call $\chi$ is really only the susceptibility above $T_c$. It is,
  nonetheless, a convenient quantity to study, and has the claimed scaling
  behavior.}

\bibitem[{\citenamefont{Ballesteros et~al.}(1996)\citenamefont{Ballesteros,
  Fernandez, Martin-Mayor, Pech, and Mu\~noz Sudupe}}]{ballesteros:96}
\bibinfo{author}{\bibfnamefont{H.~G.} \bibnamefont{Ballesteros}},
  \bibinfo{author}{\bibfnamefont{L.~A.} \bibnamefont{Fernandez}},
  \bibinfo{author}{\bibfnamefont{V.}~\bibnamefont{Martin-Mayor}},
  \bibinfo{author}{\bibfnamefont{J.}~\bibnamefont{Pech}}, \bibnamefont{and}
  \bibinfo{author}{\bibfnamefont{A.}~\bibnamefont{Mu\~noz Sudupe}},
  \emph{\bibinfo{title}{New universality class in three dimensions?: The
  antiferromagnetic $rp^2$ model}}, \bibinfo{journal}{Phys. Lett. B}
  \textbf{\bibinfo{volume}{378}}, \bibinfo{pages}{207} (\bibinfo{year}{1996}),
  \eprint{(arXiv:hep-lat/9511003)}.

\bibitem[{\citenamefont{Nightingale}(1976)}]{nightingale:76}
\bibinfo{author}{\bibfnamefont{M.~P.} \bibnamefont{Nightingale}},
  \emph{\bibinfo{title}{Scaling theory and finite systems}},
  \bibinfo{journal}{Physica A} \textbf{\bibinfo{volume}{83}},
  \bibinfo{pages}{561} (\bibinfo{year}{1976}).

\bibitem[{\citenamefont{Press et~al.}(1992)\citenamefont{Press, Teukolsky,
  Vetterling, and Flannery}}]{press:92}
\bibinfo{author}{\bibfnamefont{W.~H.} \bibnamefont{Press}},
  \bibinfo{author}{\bibfnamefont{S.~A.} \bibnamefont{Teukolsky}},
  \bibinfo{author}{\bibfnamefont{W.~T.} \bibnamefont{Vetterling}},
  \bibnamefont{and} \bibinfo{author}{\bibfnamefont{B.~P.}
  \bibnamefont{Flannery}}, \emph{\bibinfo{title}{Numerical Recipes in C, 2nd
  Ed.}} (\bibinfo{publisher}{Cambridge University Press},
  \bibinfo{address}{Cambridge}, \bibinfo{year}{1992}).

\bibitem[{\citenamefont{Ballesteros et~al.}(1998)\citenamefont{Ballesteros,
  Fern\'andez, Martin-Mayor, Mu\~noz Sudupe, Parisi, and
  Ruiz-Lorenzo}}]{ballesteros:98}
\bibinfo{author}{\bibfnamefont{H.~G.} \bibnamefont{Ballesteros}},
  \bibinfo{author}{\bibfnamefont{L.~A.} \bibnamefont{Fern\'andez}},
  \bibinfo{author}{\bibfnamefont{V.}~\bibnamefont{Martin-Mayor}},
  \bibinfo{author}{\bibfnamefont{A.}~\bibnamefont{Mu\~noz Sudupe}},
  \bibinfo{author}{\bibfnamefont{G.}~\bibnamefont{Parisi}}, \bibnamefont{and}
  \bibinfo{author}{\bibfnamefont{J.~J.} \bibnamefont{Ruiz-Lorenzo}},
  \emph{\bibinfo{title}{Critical exponents of the three-dimensional diluted
  {I}sing model}}, \bibinfo{journal}{Phys. Rev. B}
  \textbf{\bibinfo{volume}{58}}, \bibinfo{pages}{2740} (\bibinfo{year}{1998}).

\bibitem[{\citenamefont{Weigel and Janke}(2009)}]{weigel:09}
\bibinfo{author}{\bibfnamefont{M.}~\bibnamefont{Weigel}} \bibnamefont{and}
  \bibinfo{author}{\bibfnamefont{W.}~\bibnamefont{Janke}},
  \emph{\bibinfo{title}{Cross correlations in scaling analyses of phase
  transitions}}, \bibinfo{journal}{Phys. Rev. Lett.}
  \textbf{\bibinfo{volume}{102}}, \bibinfo{pages}{100601}
  (\bibinfo{year}{2009}).

\bibitem[{\citenamefont{Newman and Barkema}(1999)}]{newman:99}
\bibinfo{author}{\bibfnamefont{M.~E.~J.} \bibnamefont{Newman}}
  \bibnamefont{and} \bibinfo{author}{\bibfnamefont{G.~T.}
  \bibnamefont{Barkema}}, \emph{\bibinfo{title}{{M}onte {C}arlo Methods in
  Statistical Physics}} (\bibinfo{publisher}{Oxford University Press Inc.},
  \bibinfo{address}{New York, USA}, \bibinfo{year}{1999}).

\bibitem[{\citenamefont{Young}()}]{young:12}
\bibinfo{author}{\bibfnamefont{A.~P.} \bibnamefont{Young}},
  \emph{\bibinfo{title}{Everything you wanted to know about data analysis and
  fitting but were afraid to ask}}, \bibinfo{note}{(arXiv:1210.3781)}.

\end{thebibliography}

\end{document}